\documentclass[aip,reprint]{revtex4-1}

\draft 

\usepackage{graphicx}	
\usepackage{amsmath}	
\usepackage{amssymb}
\usepackage{bm}
\usepackage{verbatim}
\usepackage{xcolor}
\usepackage{hyperref}
\usepackage{mathtools}
\usepackage{amsfonts}
\usepackage{graphics}
\usepackage{listings}
\usepackage[normalem]{ulem}      
\usepackage{color,soul}
\begin{document}


\title{Demonstration of static atomic gravimetry using Kalman filter} 



\author{Bo-Nan Jiang}
\email[]{bnjiang@siom.ac.cn}
\affiliation{School of Science, Shenyang University of Technology, Shenyang, Liaoning 110870, China}
\affiliation{Hefei National Research Center for Physical Sciences at the Microscale and Department of Modern Physics, University of Science and Technology of China, Hefei, Anhui 230026, China}
\affiliation{Shanghai Branch, CAS Center for Excellence in Quantum Information and Quantum Physics, University of Science and Technology of China, Shanghai 201315, China}
\affiliation{Shanghai Research Center for Quantum Sciences, Shanghai 201315, China}



\setstcolor{red}

\begin{abstract}
\noindent
The measurement precision of the static atomic gravimetry is limited by white Gaussian noise in short term, which costs previous works an inevitable integration to reach the precision demanded. Here, we propose a statistical model based on the quantum projection noise and apply the associated Kalman filter to the waveform estimation in static atomic gravimetry. With the white Gaussian noise significantly removed by the the Kalman-filter formalism, the measurement noise of the gravimetry is reshaped in short term and shows $\tau^{1/2}$ feature that corresponds to random walk. During 200 hours of static measurement of gravity, the atomic gravimeter using Kalman filter demonstrates a sensitivity as good as 0.6 $\rm{nm/s^2}/\sqrt{\rm{s}}$, and highlights a precision of 1.7 $\rm{nm/s^2}$ at the measuring time of a single sample. The measurement noise achieved is also lower than the quantum projection limit below $\sim$ 30 s.
\end{abstract}


\maketitle 


\section{INTRODUCTION}
Atom interferometry realizes a versatile tool that offers precise and accurate measurement for inertial sensing\cite{Kasevich,Geiger}, which greatly complements the state of the art classical instruments\cite{Ferier,Menoret,Wu,Bidel1,Bidel2,McGuirk,Caldani,Bertoldi,Durfee,Dutta}. Among these inertial sensors, gravimeters\cite{Peters,Ferier,Menoret,Wu,Bidel1,Bidel2} are of great interest for a wide range of essential applications, from geophysics to fundamental physics\cite{Kai}.

When measuring the gravity acceleration $g$, the interferometer phase shift to be estimated is time-dependent\cite{Hu,Fu,Gillot,Ferier,Menoret}, since $g$ is a quantity that varies with time due to gravimetric Earth tides, atmospheric and polar motion effects, and other phenomena\cite{Nibauer,Torge,Rothleitner}. Meanwhile, the interferometer phase shift fluctuates due to various measurement noises that comprises vibration noise, phase noise of the Raman lasers, detection noise, and ultimately the quantum projection noise\cite{Legouet}. Experimentally, after implementing a combination of vibration compensation\cite{Hensley,Merlot,Legouet}, phase locking\cite{Santarelli,Cacciapuoti}, and efficient detection scheme\cite{Rocco}, the residual measurement noise of atomic gravimetry is dominated by white Gaussian noise in short term, and demonstrates $\tau^{-1/2}$ characteristic of reduction via integrating in time\cite{Hu,Ferier}. For static atomic gravimetry, white Gaussian noise is one of the most challenging obstacles to pursuing boosts in short-term sensitivity\cite{Legouet}, and costs previous works of precise measurements an integration time scaling from hundreds of seconds\cite{Hu,Gillot} to hours\cite{Ferier,Fu,Menoret} to reach the precision demanded.

The central statistical problem of estimating the time-varying interferometer phase shift is obviously waveform estimation\cite{Brown}. And for the waveform estimation problem of the system driven by white Gaussian processes and observed with white Gaussian noise, the Kalman filter (KF) is an optimal estimator with the minimum mean square error, and provides fast and causal estimation\cite{Kalman1,Kalman2}. Its application to atom interferometry promises to benefit applications in inertial sensing. For example, recently, the KF formalism has been experimentally implemented in a hybrid sensor to track and correct the bias drift of a classical accelerometer\cite{Cheiney}. Despite the promising expectations, a completely satisfactory implementation for static atomic gravimetry has not been found yet, since rare works shed light on the statistical model of a quantum sensor.

\begin{figure}[tbp]
\includegraphics[width=0.5\textwidth]{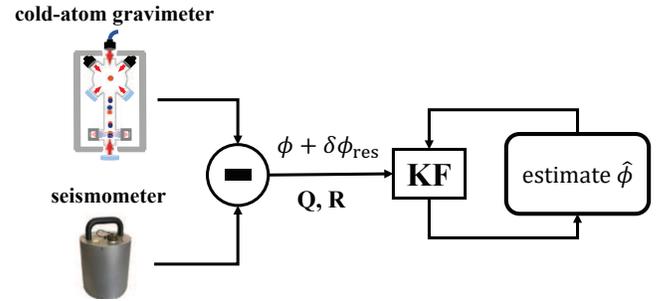}
\caption{(Color online) Schematic of the experimental setup.}
\label{fig:setup}
\end{figure}

Thus, in this work, we propose a statistical model based on the quantum projection noise and apply the associated KF formalism to the waveform estimation in static atomic gravimetry, obtaining a fast and precise estimation of the time-varying interferometer phase shift. The atomic gravimeter using KF performs the continuous gravity measurement for 200 hours at a seismic station. By comparing the KF estimates to the 725-second integration of the pre-KF signal, the reliability of the KF formalism is confirmed. With the white Gaussian noise significantly removed by the KF, the measurement noise of the gravimetry is reshaped in short term and shows $\tau^{1/2}$ feature that corresponds to random walk. The atomic gravimeter using KF demonstrates a sensitivity as good as 0.6 $\rm{nm/s^2}/\sqrt{\rm{s}}$, and highlights a precision of 1.7 $\rm{nm/s^2}$ at the measuring time of a single sample. The measurement noise achieved is also lower than the quantum projection limit below $\sim$ 30 s.

\section{EXPERIMENTAL SETUP}
\label{sec2}
The experimental setup for the atomic gravimeter using KF is presented in Fig. \ref{fig:setup}. The atomic gravimeter uses a free falling cloud of Rubidium 87 atoms as the test mass. The atoms are initially trapped and cooled by a three dimensional magneto-optical trap to 3.7 $\mu$K. After state preparation and velocity selection, $N\sim 5\times10^5$ atoms are selected and participate the interferometry that is composed of a sequence of three Raman pulses separated by two equal time intervals $T=82$ ms. After the interferometer sequence, we acquire transition probabilities from both the top ($0$) and each side ($\pm \frac{\pi}{2}$) of the central interference fringe (3 shots for each phase modulation), and determine the measured interferometer phase shift by the mid-fringe protocol\cite{Bidel3}. The direction of the momentum transfer $k_{eff}$ of Raman transitions is reversed every 9 shots in order to reject the direction-independent systematic errors. Every 5.7 s, the interferometer cycles 18 shots and outputs one interferometer phase shift $\phi_k$ that is related to the gravity acceleration $g_k$ via the scale factor
\begin{eqnarray}
\phi_k &=& k_{eff}\int^{2T}_{0}g_{\rm{s}}(t)g_ktdt \nonumber \\
     &\simeq& k_{eff}T^2g_k,
\label{eq:scale factor}
\end{eqnarray}
with $g_{\rm{s}}(t)$ being the sensitivity function of the interferometer\cite{Cheinet}. After implementing a combination of low phase noise laser system, efficient detection scheme, and seismometer correction\cite{Merlot,Legouet}, the noise on the readout of $\phi_k$ is limited by the residual measurement noise $\delta\phi_{\rm{res}}$.

The KF formalism (described below) is then applied to estimate $\phi_k$ from the readout that fluctuates due to $\delta\phi_{\rm{res}}$. The KF works in a recursive mode, that is, the previous estimate $\widehat{\phi}_{k-1}$ is used to aid in obtaining the current estimate $\widehat{\phi}_k$. In the KF loop shown in Fig. \ref{fig:setup}, the estimate $\widehat{\phi}_k$ is obtained by using the $k$th readout to update the prior estimate based on $\widehat{\phi}_{k-1}$.

As discussed in previous works\cite{Legouet,Hu,Ferier}, $\delta\phi_{\rm{res}}$ is dominated by white Gaussian noise in short term. The white Gaussian noise in readout limits the short-term sensitivity of atomic gravimeters\cite{Legouet}, and traditionally costs the gravimetry an integration time scaling from hundreds of seconds\cite{Hu,Gillot} to hours\cite{Ferier,Fu,Menoret} to retrieve the interferometer phase shift with the precision demanded.

\section{THE KALMAN FILTER}
\begin{figure*}[tbp]
\includegraphics[width=1\textwidth]{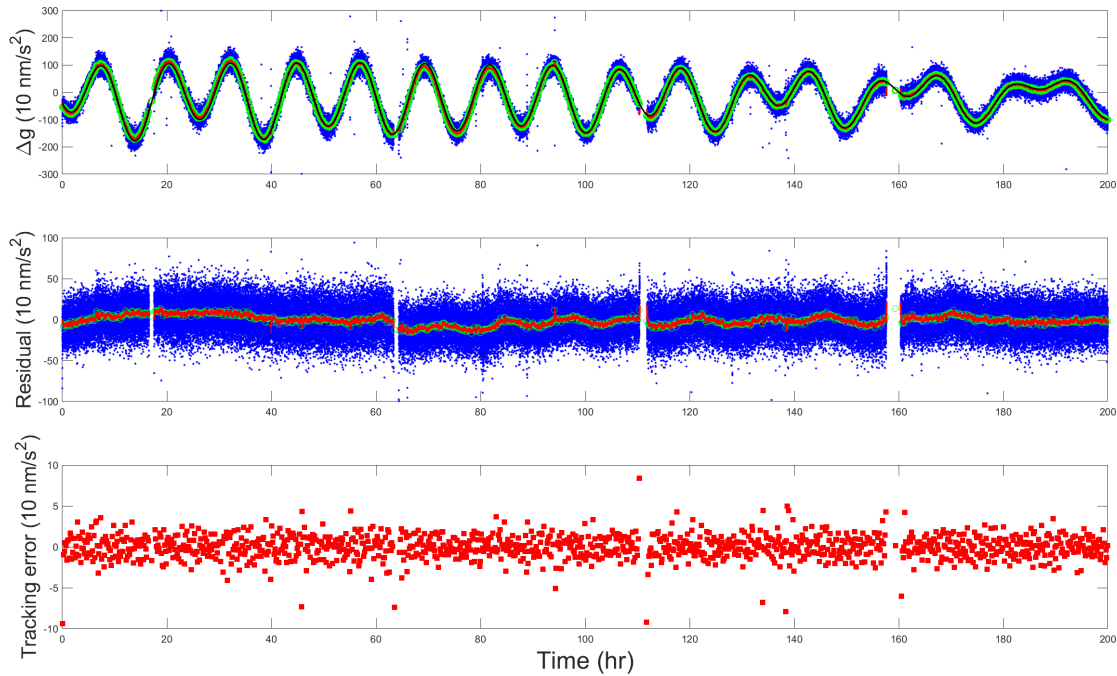}
\caption{(Color online) Static gravity measurement of 200 hours and the corresponding residual acceleration by subtracting gravimetric Earth tides from the experiment data. The tracking error (red square) is obtained by comparing the KF estimates (red) to the 725-second integration (green) of the pre-KF readout (blue). The black line is the theoretical prediction of the gravimetric Earth tides.}
\label{fig:tide}
\end{figure*}

For the linear system driven by white Gaussian processes and observed with white Gaussian noise, the KF is an optimal estimator with the minimum mean square error\cite{Kalman1,Kalman2}. The KF formalism provides a fast and causal method\cite{Brown} that reshapes the measurement noise in short term and avoids the integration. To achieve the optimal estimate, the KF relies on a statistical model for interferometer phase shift and readout dynamics.

We model the interferometer phase shift with a single-parameter state vector
\begin{equation}
  x =
  \begin{bmatrix}
  \phi
  \end{bmatrix}\!,
\label{eq:state vec}
\end{equation}
where $\phi$ is the interferometer phase shift that is related to the gravity acceleration $g$ through the scale factor $k_{eff}T^2$. Though $\phi$ is time-varying due to gravimetric Earth tides, atmospheric and polar motion effects, and other phenomena, the periods of the waveform are much longer than the time interval between two readout\cite{Nibauer,Torge,Rothleitner}. Thus, we reasonably treat the system as cyclostationary.

The discrete-time stochastic equation describing the dynamics of the interferometer phase shift is derived base on the physics fact of the cold-atom interferometry: that is, even in the cyclostationary case, the state vectors at adjacent times are not equal $x_k \neq x_{k-1}$, since the ability of resolving two different quantum phase shifts is ultimately limited by the quantum projection noise\cite{Tino}. Thus, there is always a phase error $w$ between $x_k$ and $x_{k-1}$, and $w$ stems from the quantum projection noise, which is the ultimate physical limit to atom interferometry below all the technical noise contributions\cite{Legouet}. Here, we straightforwardly describe the state by the discrete-time stochastic equation
\begin{equation}
  x_k = x_{k-1} + w_k,
\label{eq:dyn model}
\end{equation}
where $w$ describes the white Gaussian noise contribution to the state, with zero mean and variance $Q$. Here, $Q$ is the variance of the quantum projection noise and is theoretically predicted according to Ref. \cite{Tino}.

Meanwhile, the readout of the interferometer phase shift is subjected to both the residual technical noises and quantum projection noise. To account for this fact, we describe the readout by the discrete-time stochastic equation
\begin{equation}
  z_k = x_{k} + v_k,
\label{Eq:meas model}
\end{equation}
where $z_k$ is the measurement vector describing the readout process, and $v$ represents the white Gaussian noise of each readout, with zero mean and variance $R$. The variance $R$ is experimentally determined from a 10-minute sample of the pre-KF readout.

The KF works in a recursive mode. In the KF loop, the estimate $\widehat{x}_k$ and the associated error covariance matrix $P_k=E[(x_k-\widehat{x}_k)(x_k-\widehat{x}_k)^T]$ are constructed in two steps. In the propagation step, $\widehat{x}^-_k$ and $P^-_k$ are predicted conditioned on the previous posterior estimate $\widehat{x}_{k-1}$ and its covariance $P_{k-1}$ as follows
\begin{eqnarray}
\widehat{x}^-_k &=& \widehat{x}_{k-1} \nonumber \\
P^-_k &=& P_{k-1} + Q,
\label{eq:prop}
\end{eqnarray}
%


where the "super minus" indicates the prior estimate deduced according to statistical model of the system dynamics. Then, in the update step, the readout $z_k$ is used to improve $\widehat{x}^-_k$ and $P^-_k$, the posterior estimate $\widehat{x}_{k}$ and its covariance $P_{k}$ are deduced according to
\begin{eqnarray}
\widehat{x}_k &=& \widehat{x}^-_k + K_k (z_k-\widehat{x}^-_k) \nonumber \\
P_k &=& (1-K_k)P^-_k,
\label{eq:prop}
\end{eqnarray}
where the Kalman gain $K_k$ that minimizes the mean square error is computed as
\begin{equation}
  K_k = P^-_k (P^-_k + R)^{-1}.
\end{equation}
The KF loop is initialized with $\widehat{x}^-_0=z_{prior}$ and $P^-_0=Q$, according to our prior knowledge about the system. Here, $z_{prior}$ is experimentally determined by a 10-minute averaging of the pre-KF readout.

\section{PERFORMANCE}
\begin{figure}[tbp]
\includegraphics[width=0.5\textwidth]{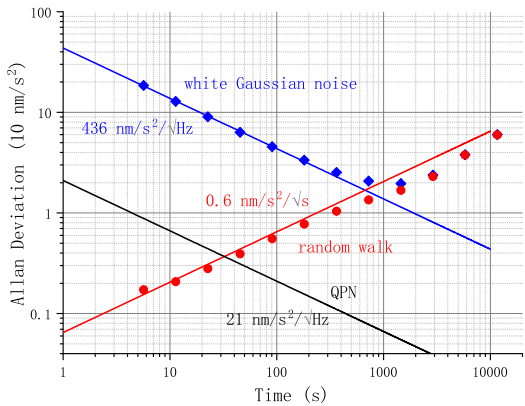}
\caption{(Color online) Allan deviation of the gravity measurement calculated from the residual acceleration of 200 hours: blue square, pre-KF gravimetry; red dot, post-KF gravimetry. The $\tau^{-1/2}$ (blue) and $\tau^{1/2}$ (red) slopes represent the corresponding averaging expected for white Gaussian noise and random walk. The black line is the sensitivity limit due to the quantum projection noise.}
\label{fig:allan}
\end{figure}

As shown in Fig. \ref{fig:tide}, at a seismic station, the experimental setup described in Sec. \ref{sec2} performed the continuous measurement of the local gravity for 200 hours, and the experiment data of the gravity acceleration were derived from the interferometer phase shift through $g\simeq\frac{\phi}{k_{eff}T^2}$. These data were chosen since they consisted of a fairly long record with sufficient quality. To test the robustness of the Kalman estimator, no preprocessing was necessary to remove gaps in the time series. The gaps are due to temporary failures of the power supply. These failures did not damage the gravimeter, and a remote restart was possible once electrical power was restored. Both the pre-KF readout (blue) and the KF estimates (red) agree well with the gravimetric Earth tides (black) predicted theoretically with an inelastic non-hydrostatic Earth model\cite{Dehant}. We also obtain the corresponding residual acceleration by subtracting the gravity signal from the theoretical prediction of the gravimetric Earth tides. The residual acceleration clearly shows that though the gaps exist in the gravimeter readings, their effects on the Kalman estimator are not significant. Note that the true gravity changes measured by the setup are more complicated than the theoretical prediction of the gravimetric Earth tides, and is usually monitored by superconducting gravimeters during the comparison\cite{WuS}. Here, without superconducting gravimeters, a 725-second integration (green) of the pre-KF readout serves as the "true" time-varing gravity signal, as $\rm{ADev}(g_{725})=20$ $\rm{nm/s^2}$ is the precision limit we can reach via the pre-KF gravimetry (see Fig. \ref{fig:allan}). To evaluate the reliability of the interferometer phase shift tracking, we compare the KF estimates directly to the 725-second integration of the pre-KF readout. The rms value of the tracking error (red square) is $\sqrt{\langle (\widehat{g}_{5.7}-g_{725})^2 \rangle}= 18$ $\rm{nm/s^2}$, which is in good agreement with the the precision limit of the pre-KF gravimetry $\rm{ADev}(g_{725})=20$ $\rm{nm/s^2}$. Obviously, the measurement noise of the pre-KF gravimetry plays a dominant role in the tracking error. And the result shows that the precision of the interferometer phase shift tracking using KF is significantly better than the precision limit of the pre-KF gravimetry.

The Allan deviation (ADev) of the residual acceleration is then calculated to characterize the short-term sensitivity and the long-term stability of the pre-KF and post-KF gravimetry. As shown in Fig. \ref{fig:allan}, the sensitivity of the pre-KF gravimetry (blue) follows 436 $ \rm{nm/s^2}/\sqrt{\rm{Hz}}$ for up to 1000 s. The measurement noise is dominated by white Gaussian noise in this regime, and demonstrates $\tau^{-1/2}$ characteristic of reduction via integrating in time. Beyond 1000 s, the stability of the pre-KF gravimetry is degraded due to phase errors that accumulate with the random walk. However, we find that the KF significantly reshapes the measurement noise by removing the white Gaussian noise in short term. The measurement noise of the post-KF gravimetry (red) then shows $\tau^{1/2}$ integrating that corresponds to random walk, and the sensitivity follows 0.6 $ \rm{nm/s^2}/\sqrt{\rm{s}}$ for up to more than $10^4$ s. Though not affecting the long-term stability, the KF does greatly improve the measurement precision in short term and highlights a precision of 1.7 $\rm{nm/s^2}$ at the measuring time of a single sample (5.7 s). Moreover, the measurement noise of the post-KF gravimetry dives under the quantum projection limit (black) below $\sim$ 30 s, which means that the KF could also be a potential method for approaching or beating the standard quantum limit in short term. Here, the sensitivity limit due to the quantum projection noise is theoretically calculated to be 21 $ \rm{nm/s^2}/\sqrt{\rm{Hz}}$ for this setup (with contrast $C=0.35$).

\section{CONCLUSION}
In conclusion, we propose a statistical model based on the quantum projection noise and apply the associated KF formalism to the waveform estimation in static atomic gravimetry, obtaining a fast and precise estimation of the time-varying interferometer phase shift. The atomic gravimeter using KF performs the continuous gravity measurement for 200 hours at a seismic station. By comparing the KF estimates to the 725-second integration of the pre-KF signal, the reliability of the KF formalism is confirmed. With the white Gaussian noise significantly removed by the KF, the measurement noise of the post-KF gravimetry is reshaped in short term and shows $\tau^{1/2}$ integrating that corresponds to random walk. The atomic gravimeter using KF demonstrates a sensitivity as good as 0.6 $\rm{nm/s^2}/\sqrt{\rm{s}}$, and highlights a precision of 1.7 $\rm{nm/s^2}$ at the measuring time of a single sample, which costs previous works an integration time scaling from hundreds of seconds to hours to reach.

The KF estimation demonstrated here increases the measurement precision of the static atomic gravimetry in short term, and makes the gravimetry "the faster, the more precise". The demonstration would be of interest for those applications involving static measurements of gravity, such as fundamental physics, geophysics, or metrology\cite{Kai}. Moreover, as this work achieves measurement noise lower than the quantum projection limit below $\sim$ 30 s, the KF could also be a potential method for approaching or beating the standard quantum limit in short term.

\section*{Acknowledgments}
We thanks Prof. Shuai Chen and his team for the work in the early stage. This work is funded by the Youth Program of National Natural Science Foundation of China (Grant No. 11804019), and also partially supported by the National Key R\&D Program of China (Grant No. 2016YFA0301601), National Natural Science Foundation of China (Grant No. 11674301), Anhui Initiative in Quantum Information Technologies (Grant No. AHY120000), and Shanghai Municipal Science and Technology Major Project (Grant No. 2019SHZDZX01).


%
%

%



\end{document}